\documentclass[11pt,oneside,onecolumn,draftcls]{IEEEtran}
\usepackage[dvips]{graphicx}
\usepackage{epsfig}
\usepackage{amsmath}
\usepackage{epsfig}
\usepackage{array}
\usepackage{amssymb}
\usepackage{color}

\begin{document}

\title{A Research Journey of Full-Duplex at University of California from Self-Interference Cancellation to Wireless Network Security}

\author{Yingbo Hua, \emph{Fellow, IEEE}
\thanks{
This work was supported in part by the Army Research Office under Grant Number W911NF-17-1-0581. The views and conclusions contained in this
document are those of the author and should not be interpreted as representing the official policies, either
expressed or implied, of the Army Research Office or the U.S. Government. The U.S. Government is
authorized to reproduce and distribute reprints for Government purposes notwithstanding any copyright
notation herein.}
\\Department of Electrical and Computer Engineering\\
University of California, Riverside, CA 92521, USA. \\
Email: yhua@ece.ucr.edu.}

%


\maketitle

\begin{abstract}
This article provides an overview of research on full-duplex at the University of California, Riverside, in the past decade. This research was initially focused on self-interference (SI) cancellation, then moved to applications of full-duplex to improve network spectral efficiency, and in recent years advanced to discover full-duplex's potentials for wireless network security. The research on SI cancellation has resulted in both hardware-based SI cancellation results and some advanced theoretical architectures which show promises but are yet to be tested via advanced hardware implementations. The applications of full-duplex for optimized spectral efficiency in ad hoc, cognitive and cellular networks have shown how to optimize power allocation among full-duplex nodes, in their antenna beamspace and over multiple subcarriers. Full-duplex has also been found to be highly beneficial for improving secrecy capacity between legitimate users against eavesdropping. Among the new capabilities that full-duplex provides for network security is an effective anti-eavesdropping channel estimation scheme which is not possible for nodes without full-duplex.
\end{abstract}


\section{Introduction}
The full-duplex research at the University of California (UC) at Riverside started in year 2010, which was inspired during the preparation of a review article on MIMO relay networks \cite{Hua_2010}. Subsequently, the author and his early collaborator Dr. P. Liang began a number of fund raising activities via industries and government agencies. We were fortunate in receiving grants from the United States Department of Defense and the UC Office of President for our research on full-duplex communications. Dr. R. Ulman of U.S. Army Research Office has continued to support our endeavor on this journey.

This article provides an overview of what we have achieved in the past ten years in the area of full-duplex. Specifically, a brief review along with some highlights will be given on the methods and their performances for self-interference cancellation required for full-duplex radio, the methods and performances of using full-duplex to improve the throughput of wireless networks, and the methods and performances of using full-duplex to improve wireless network security. In this review, we will not provide comparisons with methods developed elsewhere. But such comparisons, if available, are shown in the corresponding references listed in this paper. In some cases, the remaining challenges are also presented.

\section{Self-Interference Cancellation for Full-Duplex}
A full-duplex radio should be able to transmit and receive signals at the same time and same frequency. But the signal transmitted by the radio inevitably interferes the reception of another signal by the same radio. This is known as self-interference (SI) which is an phenomenon for all wave-based signals in nature. The fundamental task to make a full-duplex radio is to ensure that the SI is suppressed and cancelled maximally. SI cancellation is generally required after the best SI suppression has been deployed (via antenna isolation, polarization isolation, etc).

There are many methods for SI cancellation, which may involve SI cancellation via radio-frequency (RF) frontend, digital baseband, RF-baseband hybrid or any combination of these.
In \cite{Hua_Liang_2012}, the SI channel is assumed to be linear and time-invariant (LTI) so that a  linear system theory is applicable to devise a method for SI cancellation  for a MIMO SI channel. Specifically, the work in \cite{Hua_Liang_2012} proposed a time-domain transmit beamforming (TDTB) method for a multi-antenna full-duplex radio unit which has a set of receivers, a set of primary transmitters and a set of auxiliary transmitters. The primary transmitters transmit high power signals for communications with a remote node but cause the SI on the receivers.
The auxiliary transmitters apply TDTB to generate lower power auxiliary signals to cancel the SI on the receivers. The TDTB method is valid under the LTI assumption and has advantages over the frequency-domain transmit beamforming (FDTB) method. Since the Fourier transform can only be applied in practice to a signal of finite duration, the FDTB method cannot in general achieve a good SI cancellation in the prefix region of a data packet as in orthogonal frequency division multiplexing (OFDM) systems. In principle, TDTB is applicable in both baseband and RF frontend. In \cite{Hua_Liang_2012}, a hardware based experiment shows that the TDTB method reduced the SI by 50dB over a bandwidth of 30MHz.

\begin{figure}
  \centering\includegraphics[width=7cm]{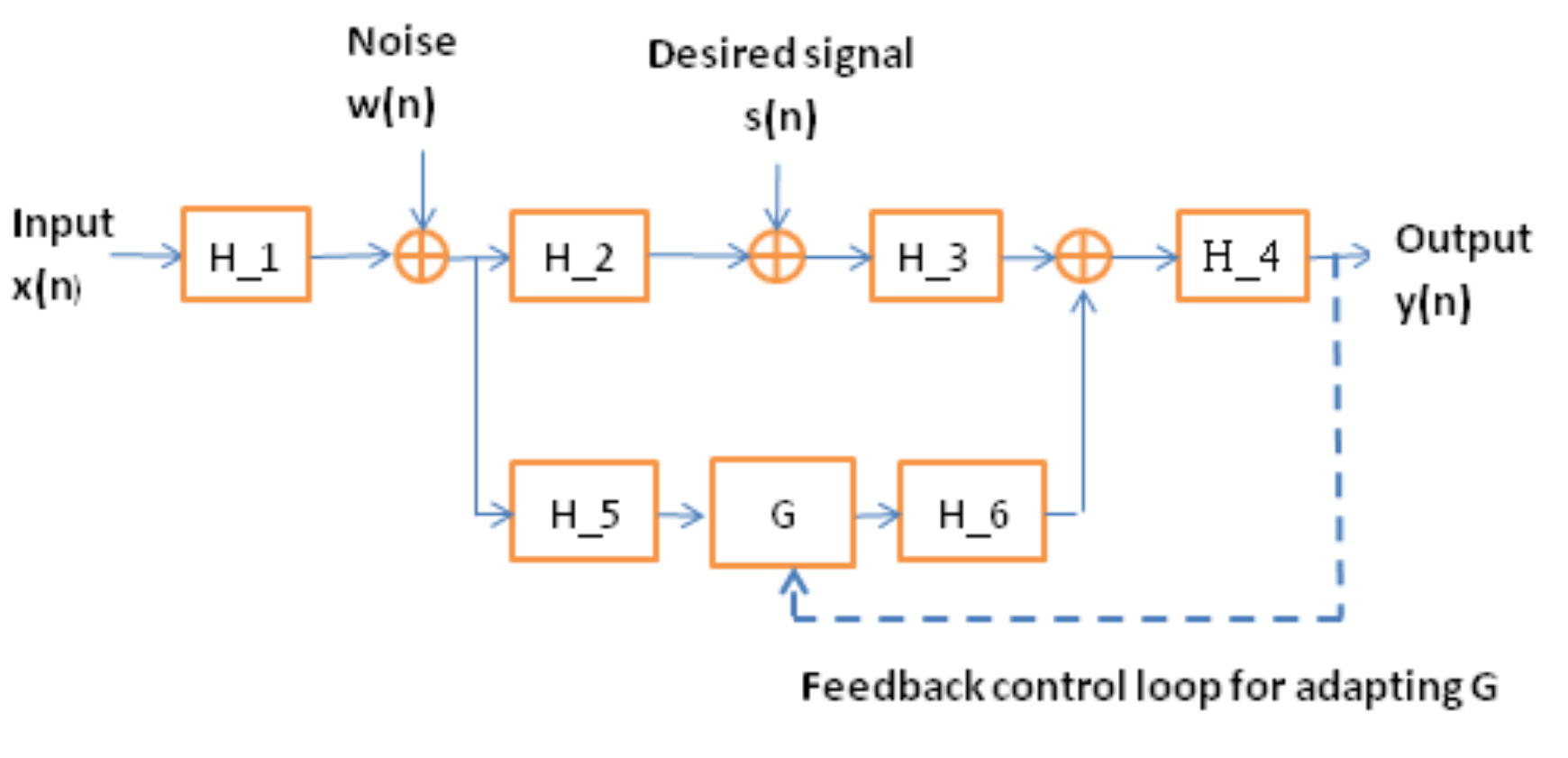}\\
  \caption{Block diagram of a SI cancellation system where the noise $w(n)$ is the transmitter noise, the desired signal $s(n)$ is from a remote node, the path of $H_5GH_6$ is the SI cancelation path, $G$ is controllable and all $H$ functions represent analog-interfaced channels or components which are (to some degree) unknown linear transfer functions \cite{Hua_2013}.  }\label{basic}
\end{figure}

The TDTB method does not address the issue of transmission noise or carrier noise that is present in any practical transmitter. Because of that, the amount of SI reduction after the use of TDTB is limited (typically 30-50dB depending on the quality of the signal generator). To overcome the SI cancellation limit caused by transmitter noise, a blind SI cancellation method was introduced in \cite{Hua_2013} where a SI cancellation chain is assumed to be directly tapped from the output of the RF power amplifier of a primary transmitter (on a full-duplex radio) and combined with the RF signal at the end of the low noise amplifier (LNA) of a receiver (on the same full-duplex radio). In this way, the noise from the transmit mixer and power amplifier no longer limits in theory the effectiveness of the SI cancellation since the transmitter noise is now a common input to both the SI channel and the SI cancellation channel (see Fig. \ref{basic}). But the challenge now is that a number of channel response functions in the SI cancellation system  are unknown (and cannot be calibrated due to analog interface), and hence the optimal parameters for the SI cancellation have to be found via blind tuning. Although some theoretical answer was given in \cite{Hua_2013} and \cite{Hua_2014}, hardware based experimental work still needs to be conducted.

\begin{figure}
 \centering\includegraphics[width=4.5cm]{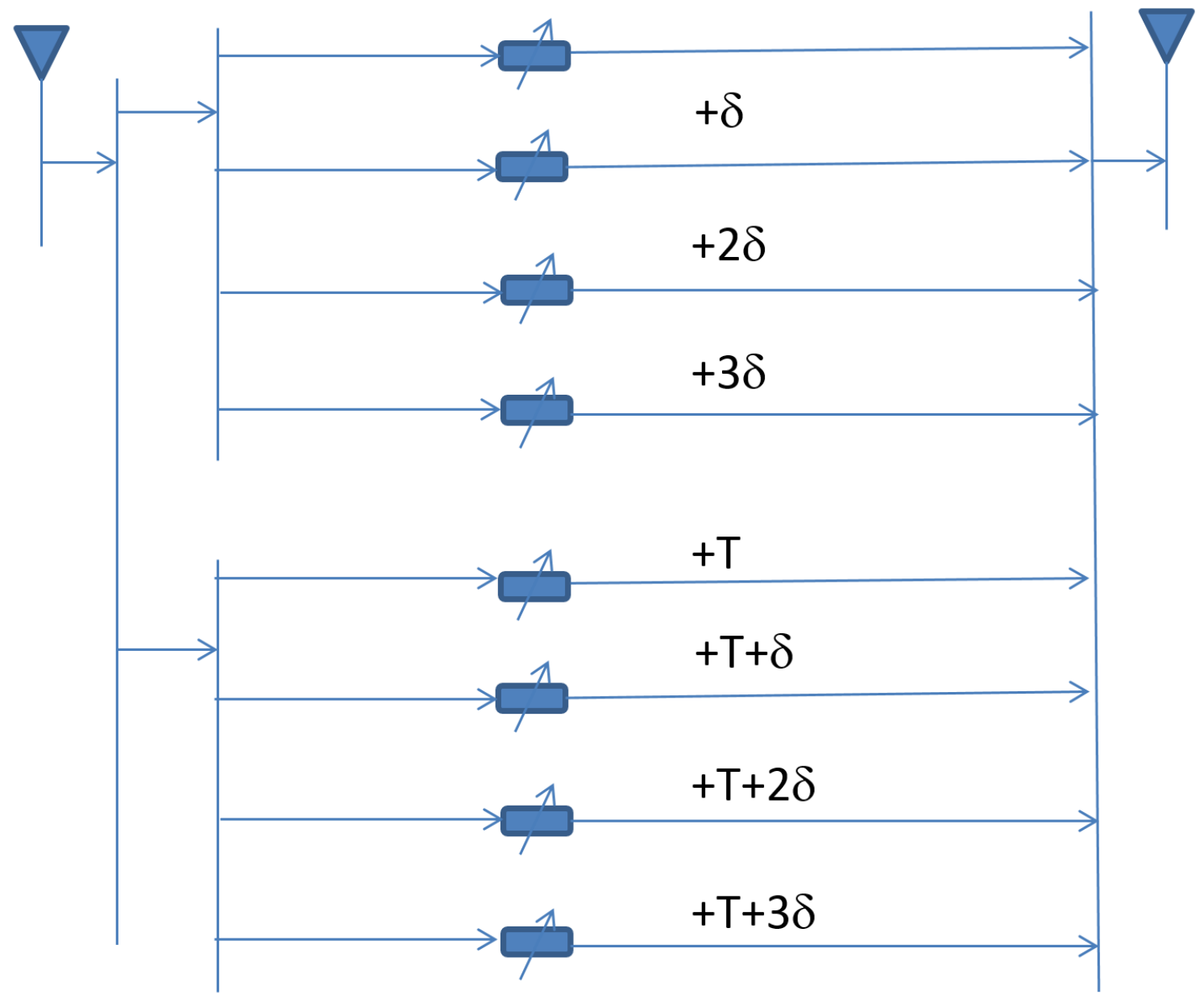}
 \caption{An all-passive SI cancelation channel architecture
 where the attenuators are clustered in groups of four, the small delay $\delta$ satisfies $f_c\delta\approx \frac{1}{4}$ with $f_c$ being the carrier frequency, and the large delay $T$ satisfies $T\ll \frac{1}{W}$ with $W$ being the bandwidth of interest \cite{Hua_2015}.
 }
 \label{fg:architecture_b}
 \end{figure}

\begin{figure}[h]
\centering\includegraphics[width=7cm]{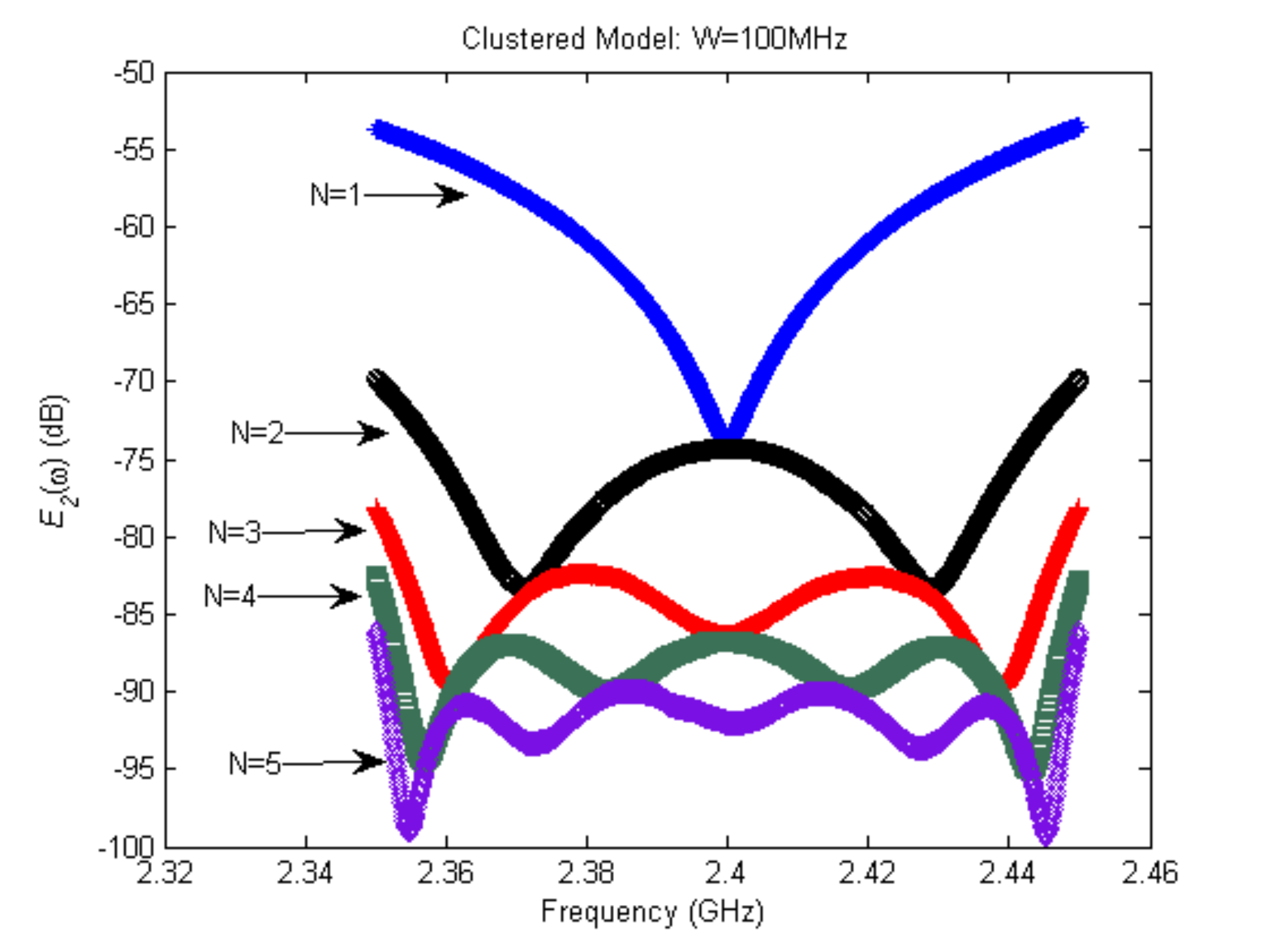}
\caption{The distribution of a normalized residual SI over frequency and $N$ after all-analog or all-passive cancellation using $T=\frac{1}{10W}$ where $N$ is the number of clusters of attenuators \cite{Hua_2015}.}
\label{fg:All_analog_over_frequency}
 \end{figure}

For the best SI cancellation, we must only use devices which have a minimum level of noise. For this reason, only passive devices should be used in the SI cancellation channel. Passive devices have much lower noise levels than active devices. One such architecture is proposed in \cite{Gholian_2014}, \cite{Hua_2014} and \cite{Hua_2015}, which is shown in Fig. \ref{fg:architecture_b}. Here all devices such as power splitters/combiners, delays and tunable attenuators are passive. Each cluster of four attenuators emulates a complex tap parameter in a transversal filter. Unlike an I/Q modulator commonly used for implementing a complex tap parameter, the clusters of attenuators are all passive devices with minimal internal noise.  Assuming that the tunable attenuators have a sufficient level of resolution, Fig. \ref{fg:All_analog_over_frequency} shows the effect of this all-passive SI cancellation chain on the residual SI. With just 5 clusters of attenuators (or total 20 attenuators), more than 90dB SI cancellation at the RF frontend is achievable over nearly 100 MHz bandwidth. (The ``all-passive'' architecture here is also referred to as ``all-analog'' architecture in \cite{Hua_2015}. The word ``passive'' refers to the nature of devices while the word ``analog'' refers to the waveforms coming into and from the devices.) Despite the promising theoretical results of the proposed architecture for SI cancellation, hardware based experiment is needed to validate the theory. One of the critical hardware requirements here is a high enough resolution of the tunable attenuators. The commonly available step-attenuators have a resolution at or above 0.5dB, which is not sufficient to achieve the results shown in Fig. \ref{fg:All_analog_over_frequency}. Another hardware requirement is about the delay devices. Achieving the delay of $T$ using a normal passive device (such as cable) can lead to something rather bulky. Hardware innovations on the tunable attenuators and delays are needed.

\section{Network Throughput Assisted by Full-Duplex}

For short range communications, the SI on a full duplex radio is much easier to manage. The existing methods based on antenna isolation along with (or even only) some baseband SI cancellation could be sufficient. In this case, applications of full-duplex radio become immediately important.

The simplest scenario of harnessing the benefit of full-duplex is a pair of single-antenna nodes where both nodes should operate in full-duplex mode if the full-duplex sum rate is higher (possibly up to a factor two) than the half-duplex sum rate,  but in half-duplex mode otherwise. The sum rate here is the sum of the data rate from node A to node B and the data rate from node B to node A. For a pair of single-antenna full-duplex nodes with any given residual SI channel gain, the (maximum) sum rate based on  either full-duplex mode or half-duplex mode can be easily found, and hence the choice of the mode is easy to make in practice.

But a more involved scenario is a pair of multi-antenna nodes. In this case, the full-duplex sum rate is not as simple as in the single-antenna case, which in fact depends on the choice of the transmission covariance matrices applied at both nodes. Furthermore, the covariance matrices should be allowed to be a function of time or have two states. In this way, the choice of the covariance matrices not only unifies but also generalizes the conventional (simplistic) notions of full-duplex mode and half-duplex mode. An optimization problem to find the covariance matrices that maximize the effective sum rate is solved in \cite{Cirik_2014}. As expected, when the residual SI level is relatively high, the optimal covariance matrices become equivalent to those in the conventional half-duplex mode. When the residual SI level is relatively low, the optimal covariance matrices become equivalent to those of two isolated channels. For an arbitrary SI level, the solution of the covariance matrices from \cite{Cirik_2014} provides the best hybrid between full-duplex and half-duplex.

Also shown in \cite{Cirik_2014}  is a multi-antenna full-duplex relay serving a multi-antenna source and a multi-antenna destination. To best harness the full-duplex capability of the relay, we let the transmission covariance matrices of the source and relay all have two states. This space-and-time freedom makes it possible to achieve the best hybrid between full-duplex and half-duplex for any given level of residual SI gain.

\begin{figure} [h]
\centering
\includegraphics[width=12.7cm]{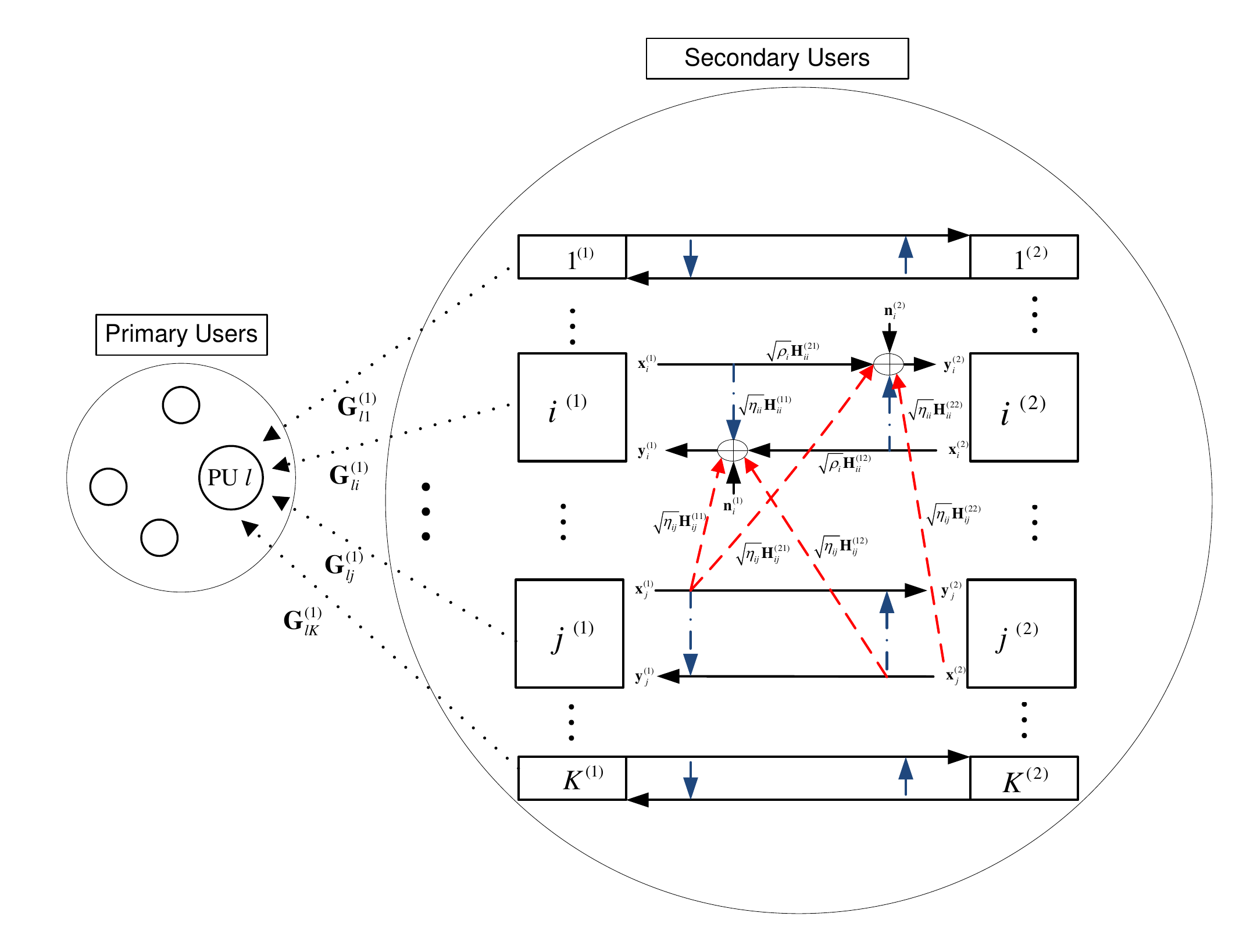}
\caption{Full-duplex MIMO links in a cognitive radio network. Squares and circles denote the SUs and PUs, respectively. The dash lines denote the interference between different pairs, the dash-and-dot lines denote the self-interference, and the dot lines denote the interference from SUs to PUs \cite{Cirik_2015_June}.}
\label{block}
\end{figure}

The space-and-time freedom embedded in the transmission covariance matrices is also known as space-time power scheduling originally proposed by the author's group in dealing with multiple links in ad hoc networks. This idea was further applied in \cite{Cirik_2015_March} and \cite{Cirik_2015_June} where multiple ($K$) links between full-duplex multi-antenna nodes are considered and the transmit covariance matrix of each node has $K$ states. In this way, any given set of the transmit covariance matrices represents a hybrid of full-duplex and half-duplex. Algorithms have been developed in \cite{Cirik_2015_March} and \cite{Cirik_2015_June} to find the optimal set of the transmit covariance matrices subject to certain types of objective functions, power constraints as well as interference constraints. Specifically, a weighted sum rate of all links is used as the objective function in \cite{Cirik_2015_March}, and the mean square errors (MSE) of the estimated symbols are used in a cost function in \cite{Cirik_2015_June}. The algorithms developed in \cite{Cirik_2015_March} and \cite{Cirik_2015_June} are also useful to mitigate the interferences from secondary users (SU) to primary users (PU) in cognitive networks as illustrated in Fig. \ref{block}.

Full-duplex radio amplifies the design space for many problems in wireless networks. Another example is a multi-carrier relay system where there may or may not be a direct link from source to destination. This is a problem addressed in \cite{Chen_2017}. These algorithmic design problems have provided a rich playground for students training in area of algorithm development.

\section{Network Security Enabled by Full-Duplex}

The original and widely known motivation for a full-duplex radio is to improve the spectral efficiency of wireless communications. But full-duplex is also quite useful to improve network security. Specifically, full-duplex can be exploited to improve secrecy against eavesdropping.

\subsection{Naive Eavesdropper}
\begin{figure}[t]
    \centering
    \includegraphics[width=6cm,draft=false]{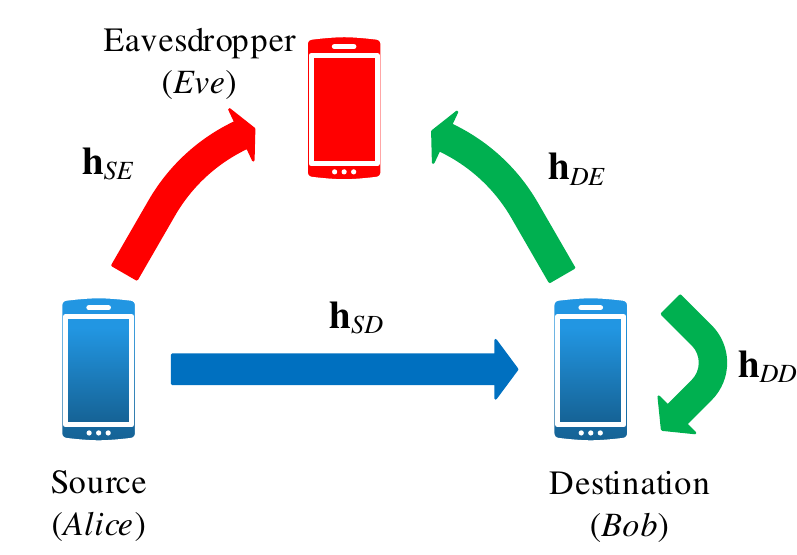}
    \caption{A three-node wireless network with a full-duplex destination \cite{Chen_2017_July}.}
    \vspace{-15pt}
    \label{Figure1}
\end{figure}

In \cite{Chen_2017_July}, a network of three nodes is considered where the first node (Alice) needs to send secret information to the second node (Bob) while the third node (Eve) is prevented from eavesdropping. If Bob is equipped with full-duplex capability, he can transmit a jamming noise to interfere Eve while he is receiving information from Alice. This is a simple setting applicable to many situations of short range communications such as Bluetooth communications between friends in the presence of a stranger as illustrated in Fig. \ref{Figure1}. Assuming that multiple subcarriers are used and all channel state information is known to Alice and Bob, they can both apply optimal power allocation over the subcarriers to maximize the secrecy capacity (in bits per second per Hertz) from Alice to Bob against Eve. The basic idea behind this work is as follows.  For different subcarriers, the channel strength from Alice to Bob may be better or worse than that from Alice to Eve, which is random due to multipath effect. The transmit power allocation at Alice should favor the subcarriers that are relatively strong for the legitimate channel or equivalently relatively weak for the eavesdropping channel. The residual SI channel gain at Bob is also random from subcarrier to subcarrier due to multipath. If the residual SI channel for a subcarrier is weak compared to that from Bob to Eve, Bob should apply more jamming power on that subcarrier to jam Eve without affecting himself much. Because of the large number of variables in the system, only an efficient optimization algorithm can quickly yield the best power scheduling in real time, which is done in \cite{Chen_2017_July}.

\subsection{Covert Eavesdropper}
The eavesdropper considered in \cite{Chen_2017_July} is naive and non-covert. But for many applications such as in military, we need to handle covert eavesdroppers whose locations and channel conditions are generally unknown to legitimate users.

\begin{figure}
  \centering
  \includegraphics[width=6cm]{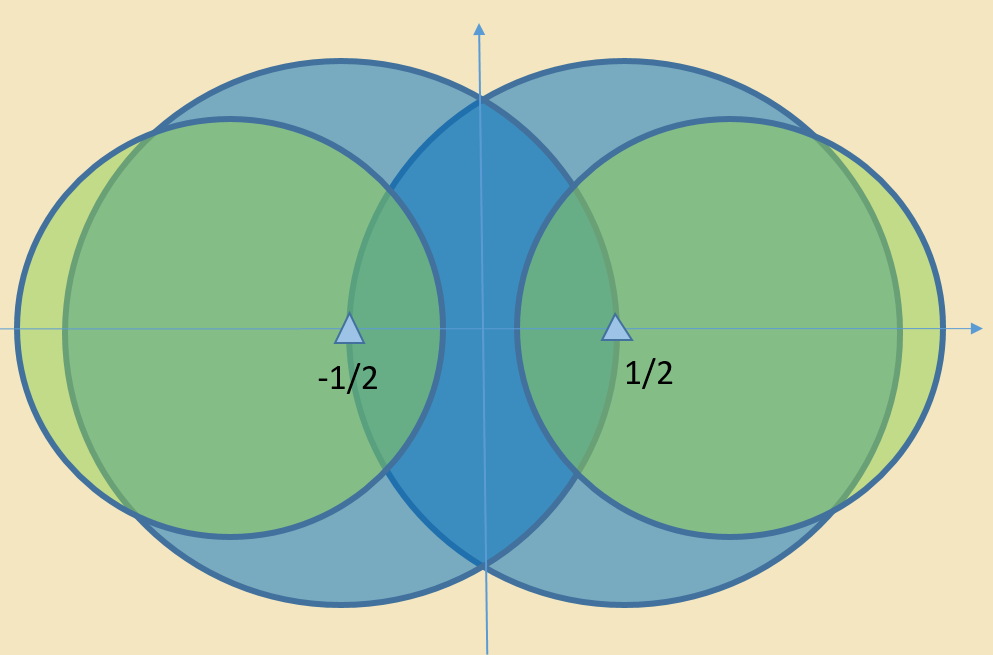}\\
  \caption{``There is a jamming power $P_J\geq 0$ such that the secrecy capacity $S_{x,y}$ against Eve at location $(x,y)$ is positive'' if and only if Eve is not in the dark blue region. This dark blue region vanishes if and only if the normalized SI channel gain $\rho$ is less than one \cite{Hua_2017}.}\label{almond_with_partial_removal}
\end{figure}

In \cite{Hua_2017}, we have considered a pair of full-duplex single-antenna users (Alice and Bob) against a single-antenna eavesdropper (Eve) located anywhere in a two-dimensional space. Assuming a fixed relationship between the channel gain (between any two nodes) and the distance (between the two nodes), the work in \cite{Hua_2017} provides geometrical regions of Eve against which secrecy capacity may be positive or zero. For example, Fig. \ref{almond_with_partial_removal} illustrates such regions. Here the secrecy capacity $S_{x,y}$ is the average of the secrecy capacity for transmission from Alice at location $(-\frac{1}{2},0)$ to Bob at location $(\frac{1}{2},0)$ and the secrecy capacity for transmission from Bob to Alice, and both transmissions are against Eve at the location $(x,y)$. And $P_J$ is the jamming power applied by Alice or Bob when operating in full-duplex mode, and $\rho$ is a normalized residual SI channel gain. As long as $\rho<1$, there will be a positive secrecy capacity against Eve located anywhere. It is important to note that for any given residual SI channel gain of a full-duplex radio, $\rho$ is a decreasing function of the distance between the two legitimate users. Therefore, as long as the range of communications between Alice and Bob is upper bounded by some value (governed by an actual residual SI channel gain), the two full-duplex nodes can exchange information with each other with a positive secrecy capacity against Eve in any location.

The results in \cite{Hua_2017} are based on single-antenna Eve, which exclude the situations where many covert eavesdroppers might collude with each other to form a virtual antenna array at the physical layer. This is a challenging problem. In \cite{Hua_2018}, we have shown that regardless of how two full-duplex users exploit their full-duplex capability in the conventional ways, the secrecy capacity quickly vanishes if the number of antennas on Eve increases. This phenomenon also holds for multi-antenna full-duplex users.

For strong security, it is desirable to develop transmission schemes that have a significant positive secrecy unconditional on the number of antennas on Eve. This unconditional secrecy (UNS) is fortunately achievable with a novel use of full-duplex. In  \cite{Hua_2018}, a method called anti-eavesdropping channel estimation (ANECE) is proposed, which  allows two or more full-duplex legitimate users (such as legitimate drones) to estimate their own channel state information (CSI) with respect to each other but prevents any Eve (with any number of antennas) from obtaining any consistent estimate of its CSI with respect to any user. It is shown in \cite{Hua_2018} and \cite{Sohrabi_2019} that for some limited transmission time per CSI coherence period, ANECE yields a significant positive UNS. If the number of antennas on each user is not large, an efficient, reliable and secure transmission between users may require coding over multiple CSI coherence periods.


A core idea behind ANECE is a special type of pilot signals concurrently transmitted by the full-duplex users. The concurrency is at the level of symbol interval, which (along with clock errors) sets some limit on how far apart the legitimate users should be from each other. The pilot signals are such that the receive channel matrix at any user is always observable by the user but a subspace of the receive channel matrix at any Eve is unobservable by the Eve. Subject to the above requirement, there exist many choices of the pilots for ANECE. But the optimal designs of the pilots for ANECE have been addressed recently in \cite{Zhu_2020}. The difference between an initially suggested pilot and an optimal pilot can be up to 20dB in terms a normalized mean squared error (MSE) of the estimated channel matrices for users (see Fig. 2 in \cite{Zhu_2020}).

\section{Conclusion}
For full-duplex communications, self-interference (SI) isolation and cancellation are the most fundamental issue. While progress has been made at UC Riverside and elsewhere, and full-duplex is becoming feasible for short range communications, much more progress is still needed especially in hardware devices and implementations in order to increase the range of full-duplex communications. We hope that the theoretical architectures highlighted earlier for SI cancellation
 could help motivate a good investment in the required hardware development. The applications of full-duplex to improve network-wise spectral efficiency  are still worthy directions of pursuit for graduate students but more importantly for industrial network designing and planning where full-duplex is considered as an option. The recent discovery of full-duplex's potentials for wireless network security should drive further  investigations. This remains an active topic of research at UC Riverside.


\begin{thebibliography}{99}

\bibitem{Hua_2010} Y. Hua, ``An overview of beamforming and power allocation for MIMO relays,'' Proc. of MILCOM, pp. 99-104, San Jose, CA, Nov. 2010.

\bibitem{Hua_Liang_2012} Y. Hua, P. Liang, Y. Ma, A. Cirik, Q. Gao, ``A method for broadband full-duplex MIMO radio,'' IEEE Signal Processing Letters, Vol. 19, No. 12, pp. 793-796, Dec 2012.

\bibitem{Hua_2013} Y. Hua, Y. Ma, P. Liang, A. Cirik, ``Breaking the barrier of transmission noise in full-duplex radio'', Proc. of MILCOM, pp. 1558-1563, San Diego, CA, Nov 2013.

\bibitem{Gholian_2014} A. Gholian, Y. Ma, Y. Hua, ``A numerical investigation of all-analog radio self-interference cancellation,'' Proc. of IEEE Workshop on SPAWC, pp. 459-463, Toronto, Canada, June 2014.

\bibitem{Hua_2014} Y. Hua, Y. Li, C. Mauskar, Q. Zhu, ``Blind digital tuning for interference cancellation in full-duplex radio'', Proc. of Asilomar Conference on Signals, Systems and Computers, pp. 1691-1695, Pacific Grove, CA, Nov 2014

\bibitem{Hua_2015} Y. Hua, Y. Ma, A. Gholian, Y. Li, A. Cirik, P. Liang, ``Radio self-interference cancellation by transmit beamforming, all-analog cancellation and blind digital tuning,'' Signal Processing, Vol. 108, pp. 322-340, 2015.

    \bibitem{Cirik_2014} A. Cirik, Y. Rong, Y. Hua, ``Achievable rates and QoS considerations of full-duplex MIMO radios for fast fading channels with imperfect channel estimation,'' IEEE Transactions on Signal Processing, Vol. 62, No. 15, pp. 3874-3886, Aug 2014.

\bibitem{Cirik_2015_March} A. Cirik, R. Wang, Y. Hua, M. Latva-Aho, ``Weighted sum-rate maximization for full-duplex MIMO interference channels,'' IEEE Transactions on Communications, Vol. 63, No. 3, pp. 801-815, March 2015.

\bibitem{Cirik_2015_June} A. Cirik, R. Wang, Y. Rong, Y. Hua, ``MSE based transceiver designs for full-duplex MIMO cognitive radios,'' IEEE Transactions on Communications, Vol. 63, No. 6, pp. 2056-2070, June 2015.

\bibitem{Chen_2017} L. Chen, W. Meng, Y. Hua, ``Optimal power allocation for a full-duplex multicarrier decode-forward relay system with or without direct link'', Signal Processing, 137 (2017) 177-191.

\bibitem{Chen_2017_July} L. Chen, Q. Zhu, W. Meng, Y. Hua, ``Fast power allocation for secure communication with full-duplex radio'', IEEE Transactions on Signal Processing, Vol. 65, No. 14, pp. 3846-3861, July 2017.

 \bibitem{Hua_2017}   Y. Hua, Q. Zhu, R. Sohrabi, ``Fundamental properties of full-duplex radio for secure wireless communications,'' http://arxiv.org/abs/1711.10001, pp. 1-13, 2017.

\bibitem{Hua_2018}
Y. Hua, ``Advanced properties of full-duplex radio for securing wireless network,'' IEEE Transactions on Signal Processing, Vol. 67, No. 1, pp. 120-135, Jan. 2019.

\bibitem{Sohrabi_2019}
R. Sohrabi, Q. Zhu,  Y. Hua, ``Secrecy analyses of a full-duplex MIMOME network,'' IEEE Transactions on Signal Processing, Vol. 67, No. 23, pp. 5968-5982, Dec. 2019.

\bibitem{Zhu_2020}
Q. Zhu, S. Wu, Y. Hua, ``Optimal pilots for anti-eavesdropping channel estimation,'' IEEE Transactions on Signal Processing, Vol. 68, pp. 2629-2644, 2020.


\end{thebibliography}
\end{document}